\documentclass[aps,pra]{revtex4}

\usepackage{amssymb}
\usepackage{graphicx}
\usepackage{dcolumn}
\usepackage{bm}
\usepackage{amsmath}
\usepackage{epsfig}

\begin{document}

\title{Thermodynamics of bipartite systems:\\Application to light-matter
interactions}

\author{E. Boukobza}
\author{D.J. Tannor}
\affiliation{Department of Chemical Physics,\\
Weizmann Institute of Science, Rehovot 76100, Israel}

\begin{abstract}

Heat and work for quantum systems governed by dissipative master
equations with a time-dependent driving field were introduced in the
pioneering work of Alicki [J. Phys. A {\bf12}, L103 (1979)].
Alicki's work was in the Schr\"{o}dinger picture; here we extend
these definitions to the Heisenberg and interaction pictures. We
show that in order to avoid consistency problems, the full time
derivatives in the definitions for heat flux and power (work flux)
should be replaced by partial time derivatives. We also present an
alternative approach to the partitioning of the energy flux which
differs from that of Alicki in that the instantaneous interaction
energy with the external field is not included directly. We then
proceed to generalize Alicki's definition of power by replacing the
original system and its external driving field with a larger,
bipartite system, governed by a time-\textit{independent}
Hamiltonian. Using the definition of heat flux and the generalized
definition of power, we derive the first law of thermodynamics in
differential form, both for the full bipartite system and the
partially traced subsystems. Although the second law (Clausius
formulation) is satisfied for the full bipartite system, we find
that in general there is no rigorous formulation of the second law
for the partially traced subsystem unless certain additional
requirements are met. Once these requirements are satisfied,
however, both the Carnot and the Clausius formulations of the second
law are satisfied. We illustrate this thermodynamic analysis on both
the simple Jaynes-Cummings model (JCM) and an extended dissipative
Jaynes-Cummings model (ED-JCM), which is a model for a quantum
amplifier.

\end{abstract}

\maketitle

\section{Introduction\label{sec:Intro}}

Thermodynamics of quantum systems has intrigued scientists since the
early years of the development of quantum mechanics. von Neumann
\cite{von Neumann} defined an entropy function which is based on the
density matrix, and is similar in spirit to the Gibbs entropy. It
takes into account both the populations and the coherences of the
density matrix. Born \cite{Born} partitioned the energy of various
types of quantum statistical systems into heat and work. Tolman
\cite{Tolman} defined an entropy function which is identical with
the Gibbs entropy, and which is based on a quantum version of
Boltzmann's $H$-function. He showed that Carnot's formula holds for
a quantum canonical distribution, and that Clausius's inequality
holds in general. Scovil and Schulz-DuBois \cite{Scovil01} analyzed
a three-level quantum system coupled to two thermal reservoirs. Much
later, a master equation for open quantum systems in the weak
coupling Markovian regime was developed by Lindblad, who showed that
under a trace-preserving completely positive map the relative
entropy is non-increasing \cite{Lindblad2}. Spohn defined the
entropy production function, and showed it is always positive
\cite{Spohn}. Pusz and Wornowicz defined work in a general $C^{*}$
algebraic context for systems with varying external forces
\cite{Pusz}.  Based on the Markovian master equation, Alicki
\cite{Alicki} defined work and heat for systems with time dependent
Hamiltonians. The ideas of Alicki were implemented in a series of
papers by Kosloff and coworkers on quantum heat engines
\cite{Kosloff01}, \cite{Geva01}, \cite{Kosloff02}.

This work continues the work of Alicki. The main focus of this paper
is laying the foundations for a complete thermodynamic analysis of
bipartite systems, and to apply this novel approach to quantum
optical systems. It is not obvious how to generalize the
thermodynamic definition of Alicki to quantum optical systems, since
Alicki's definition of work presupposes a time-dependent external
field, whereas in quantum optical systems the Hamiltonian is time-independent.
To solve this problem, we begin by noting that the
time-dependence of the Hamiltonian in the semiclassical treatment
arises from the time-independent Hamiltonian in the quantum system
because we are looking at only a part of the larger system. For
example, in the absence of dissipation the total energy of the
bipartite matter-light system is conserved, but the energy of each
subsystem in general is time-dependent. If one observes the energy
changing with time in one of the subsystems, there is no way of
distinguishing whether this is as a result of external forcing or
because the subsystem is part of the larger bipartite system.
Therefore, if work can be defined for the former we can expect that
it can be defined for the latter as well. The central result of this
paper, eq.  \ref{WAA}, provides precisely such an expression for the
work, expressed in terms of the Hamiltonian of the full bipartite
system. Inspection of this equation shows that it is not restricted
to quantum optical systems, and is in fact completely general,
applying to any bipartite system.  Nevertheless, the application to
quantum optical systems is particularly intriguing since it lays the
foundation for a thermodynamic analysis of light-matter interactions
in non-equilibrium systems.

In Section \ref{sec:Alicki}, we review Alicki's definition of
quantum heat and work. In Section \ref{sec:Reps} we present a minor
generalization of Alicki's definitions for unipartite systems that
provides thermodynamic consistency in the Schr\"{o}dinger, the
Heisenberg, and the interaction pictures. In section \ref{sec:Alt}
we suggest an alternative partitioning of energy into heat and work
that differs from that of Alicki and presages the bipartite
formulation in Section \ref{sec:Bipartite}. In Section
\ref{sec:SecondUni} we analyze the different formulations of the
second law for unipartite systems. Section \ref{sec:Bipartite}, on
bipartite systems, is the main part of this paper. We begin with the
derivation of the general formula for work flux (power) and heat
flux in bipartite systems that generalizes our unipartite
formulation of the first law in Section \ref{sec:Alt}. In section
\ref{sec:Bipartiteb} we show that although the second law in the
Clausius formulation is satisfied for the full bipartite system, in
general there is no rigorous formulation of the second law for the
partially traced subsystem unless certain additional requirements
are met. Once these requirements are satisfied, however, both the
Carnot and the Clausius formulations of the second law are
satisfied. In Section \ref{sec:Apps} we apply the general formalism
to the simple Jaynes-Cummings model (JCM) and to an extended
dissipative JCM, which serves as a quantum amplifier. Section
\ref{sec:Conc} concludes.

\section{Unipartite Systems\label{sec:Unipartite}}
\subsection{Energy flux of systems governed by master equations\label{sec:Alicki}}

The master equation for a system coupled to thermal reservoirs in
the Schr\"{o}dinger picture and the Markovian regime is given by:
\begin{equation}
\dot{\rho}=\mathcal{L}[\rho]=\mathcal{L}_{h}[\rho]+\mathcal{L}_{d}[\rho],\label{Master}
\end{equation}
where $\mathcal{L}[\rho]$ is a general Lindblad super operator,
$\mathcal{L}_{h}[\rho]=-\frac{i}{\hbar}[H(t),\rho]$ is the
Hamiltonian Liouville super operator, and $\mathcal{L}_{d}[\rho]$ is
a dissipative Lindblad super operator. Based on the master equation
one can calculate the average energy by:
\begin{equation}
\langle E\rangle=\textrm{Tr}\{\rho(t)H(t)\}\label{energy}.
\end{equation}
Alicki \cite{Alicki} partitioned the energy into \textit{heat} and
\textit{work} by taking the time derivative of equation
\ref{energy}. Heat is defined by:
\begin{equation}
Q\equiv\int\limits_{0}^{t}
\textrm{Tr}\left\{\frac{d\rho(t')}{dt'}H(t')\right\}dt',\label{heatAL}
\end{equation}
where work was defined originally by Pusz and Wornowicz \cite{Pusz}
as:
\begin{equation}
W\equiv\int\limits_{0}^{t}
\textrm{Tr}\left\{\rho(t')\frac{dH(t')}{dt'}\right\}dt'.\label{workAL}
\end{equation}
A consistency problem arises when one applies these definitions in
the interaction picture, and one has to be careful when defining a
Heisenberg picture for a system governed by dissipative evolution.
Moreover, according to eq. \ref{workAL} a system whose interaction
with some external degree of freedom is governed by a time
\textit{independent} Hamiltonian does not perform any work (as in
the case of quantum electrodynamics where field-matter interactions
are governed by time independent Hamiltonians). These two issues
will be addressed in the sections to follow this brief introduction.

\subsection{Heat and work in various pictures for unipartite systems
coupled to a time dependent external field\label{sec:Reps}}
\subsubsection{The Schr\"{o}dinger picture\label{sec:Shroedinger}}

The time evolution of the average value of an operator is given by:
\begin{equation}
\frac{d\langle
A^{S}\rangle}{dt}=\frac{d(\textrm{Tr}\{\rho^{S}A^{S}\})}{dt}=\textrm{Tr}\left\{\frac{d\rho^{S}}{dt}A^{S}\right\}+\textrm{Tr}\left\{\rho^{S}\frac{dA^{S}}{dt}\right\}
=\textrm{Tr}\left\{\frac{\partial\rho^{S}}{\partial
t}A^{S}\right\}+\textrm{Tr}\left\{\rho^{S}\frac{\partial
A^{S}}{\partial t}\right\},\label{Sch}
\end{equation}
where in the Schr\"{o}dinger picture
$d\rho^{S}/dt=\partial\rho^{S}/\partial t$ and $dA^{S}/dt=\partial
A^{S}/\partial t$. Consider a system whose evolution is governed by
equation \ref{Master}. The first term in eq. \ref{Sch} becomes:
\begin{equation}
\textrm{Tr}\left\{\frac{\partial\rho^{S}}{\partial
t}A^{S}\}=\textrm{Tr}\{\mathcal{L}[\rho^{S}]A^{S}\}=-\frac{i}{\hbar}\textrm{Tr}\{[H^{S},\rho^{S}]A^{S}\}+
\textrm{Tr}\{\mathcal{L}_{d}[\rho^{S}]A^{S}\right\}.
\end{equation}
Using the definition for quantum heat (eq. \ref{heatAL}) the heat
flux in the Schr\"{o}dinger picture is given by:
\begin{equation}
\dot{Q}=\textrm{Tr}\left\{\frac{d\rho^{S}}{dt}H^{S}\right\}=\textrm{Tr}\left\{\frac{\partial\rho^{S}}{\partial
t}H^{S}\right\}=\textrm{Tr}\{\mathcal{L}_{d}[\rho^{S}]H^{S}\},\label{QSch}
\end{equation}
where in the last equality we have used the fact that
$\textrm{Tr}\{\mathcal{L}_{h}[\rho^{S}]H^{S}\}=0$. The power (work
flux) is thus given by:
\begin{equation}
P\equiv\dot{W}=\textrm{Tr}\left\{\rho^{S}\frac{dH^{S}}{dt}\right\}=\textrm{Tr}\left\{\rho^{S}\frac{\partial
H^{S}}{\partial t}\right\}.\label{PSch}
\end{equation}
Note that in case of only Hamiltonian dynamics there is no heat
involved, and if the Hamiltonian is time independent there is no
work done by the system.

\subsubsection{The Heisenberg picture\label{sec:Heisenberg}}

Consider a system whose evolution is purely Hamiltonian:
\begin{equation}
\dot{\rho^{S}}=\mathcal{L}_{h}[\rho^{S}]=-\frac{i}{\hbar}[H^{S},\rho^{S}].\label{SchH}
\end{equation}
Eq. \ref{SchH} can be formally integrated:
\begin{equation}
\rho^{S}(t)=U(t)\rho(0)U^{\dag}(t),\label{SchHint}
\end{equation}
where $U(t)$ is a unitary operator. If the Hamiltonian is time
independent then $U(t)=\exp^{-\frac{i}{\hbar}H^{S}t}$. Otherwise,
$U(t)$ is obtained by the time ordering procedure \cite{Mukamel}.
The essence of the Heisenberg representation is that operators
'move' in time, and it is defined by \cite{Mukamel}:
\begin{equation}
A^{H}\equiv U^{\dag}(t)A^{S}U(t),\label{HeisO}
\end{equation}
where $U(t)$ is the same unitary operator that appears in eq.
\ref{SchHint}. Note that in this case $\rho^{H}$ is \textit{time
independent}:
\begin{equation}
\dot{\rho^{H}}=\frac{i}{\hbar}U^{\dag}[H^{S},\rho^{S}]U+U^{\dag}\dot{\rho^{S}}U=\frac{i}{\hbar}([H^{H},\rho^{H}]-[H^{H},\rho^{H}])=0
\end{equation}
Note also that if $H^{S}$ is time independent
$H^{H}(t)=H^{S}(t)=H^{S}(0)$. The average value of an operator is
independent of representation:
\begin{equation}
\langle
A^{S}\rangle=\textrm{Tr}\{\rho^{S}A^{S}\}=\textrm{Tr}\{U^{\dag}(t)\rho^{S}U(t)U^{\dag}(t)A^{S}U(t)\}=\textrm{Tr}\{\rho^{H}A^{H}\}=\langle
A^{H}\rangle.
\end{equation}

Consider now a system whose evolution is governed by eq.
\ref{Master}. That is, its evolution is also dissipative. A solution
in the form of eq. \ref{SchHint} can not be obtained due to the
non-unitary nature of $\mathcal{L}_{d}[\rho]$. In this case
$\rho^{H}$ is \textit{time dependent}:
\begin{equation}
\dot{\rho^{H}}=\frac{i}{\hbar}U^{\dag}[H^{S},\rho^{S}]U+U^{\dag}\dot{\rho^{S}}U=U^{\dag}\mathcal{L}_{d}[\rho^{S}]U\equiv\mathcal{L}_{d}^{H}[\rho^{H}].\label{Heisrho}
\end{equation}
The evolution of the density operator in the Heisenberg picture is
due to the dissipative part in the master equation, while the
Hamiltonian evolution is still canceled out. The time dependence of
$\rho^{H}$ is not an artifact of a definition or a transformation.
Consider any function of the density matrix, for example purity
($\langle \rho\rangle=\textrm{Tr}\{\rho^{2}\}$). It is obvious that
under unitary evolution, purity does not change with time (as can be
shown by a Taylor expansion of the density operator). However, under
dissipative dynamics purity may change with time, and this should be
the case in any physical picture. Therefore, when one calculates the
average value of an operator there are two contributions to its
evolution: the 'moving' density operator, and the 'moving'
observable. This can be seen by:
\begin{equation}
\frac{d\langle
A^{H}\rangle}{dt}=\frac{d(\textrm{Tr}\{\rho^{H}A^{H}\})}{dt}=\textrm{Tr}\left\{\frac{d\rho^{H}}{dt}A^{H}\right\}+\textrm{Tr}\left\{\rho^{H}\frac{dA^{H}}{dt}\right\}.\label{Heisavt}
\end{equation}
The first term in eq. \ref{Heisavt} is obtained by substitution of
eq. \ref{Heisrho}:
\begin{equation}
\textrm{Tr}\left\{\frac{d\rho^{H}}{dt}A^{H}\right\}=\textrm{Tr}\{\mathcal{L}_{d}^{H}[\rho^{H}]A^{H}\},\label{preheatflux}
\end{equation}
while the second term is given by:
\begin{equation}
\textrm{Tr}\left\{\rho^{H}\frac{dA^{H}}{dt}\right\}=\frac{i}{\hbar}\textrm{Tr}\{\rho^{H}[H^{H},
A^{H}]\}+\textrm{Tr}\left\{\rho^{H}\left(\frac{\partial A}{\partial
t}\right)^{H}\right\}.\label{prepower}
\end{equation}
Substituting $H^{H}$ in eq. \ref{preheatflux} and eq. \ref{prepower}
shows that the definitions for heat flux and power in the Heisenberg
picture are identical to those in the Schr\"{o}dinger picture:
\begin{eqnarray}
\textrm{Tr}\left\{\frac{d\rho^{H}}{dt}H^{H}\right\}&=&\textrm{Tr}\{\mathcal{L}_{d}^{H}[\rho^{H}]H^{H}\}=\textrm{Tr}\{\mathcal{L}_{d}^{S}[\rho^{S}]H^{S}\}=\dot{Q}\\
\textrm{Tr}\left\{\rho^{H}\frac{dH^{H}}{dt}\right\}&=&\textrm{Tr}\left\{\rho^{H}\left(\frac{\partial
H}{\partial
t}\right)^{H}\right\}=\textrm{Tr}\left\{\rho^{S}\left(\frac{\partial
H}{\partial t}\right)^{S}\right\}=P.
\end{eqnarray}

Geva and Kosloff rewrite eq. \ref{Sch} using the cyclic invariance
of the trace. By doing this they obtain an alternative Heisenberg
picture in which the observable is evolved by the dissipative part
of the master equation together with the Hamiltonian part
\cite{Geva00}, \cite{Kos00}:
\begin{eqnarray}
\frac{d\langle
A^{S}\rangle}{dt}&=&\textrm{Tr}\left\{\frac{\partial\rho^{S}}{\partial
t}A^{S}\right\}+\textrm{Tr}\left\{\rho^{S}\frac{\partial
A^{S}}{\partial
t}\right\}=\textrm{Tr}\{\mathcal{L}[\rho^{S}]A^{S}\}+\textrm{Tr}\left\{\rho^{S}\frac{\partial
A^{S}}{\partial
t}\right\}\\\nonumber&=&\textrm{Tr}\{\rho^{S}\mathcal{L}^{\star}[A^{S}]\}+\textrm{Tr}\left\{\rho^{S}\frac{\partial
A^{S}}{\partial
t}\right\}=\textrm{Tr}\left\{\rho^{S}\frac{dA}{dt}\right\},\label{KosGev}
\end{eqnarray}
where:
\begin{equation}
\frac{dA}{dt}\equiv\mathcal{L}^{*}[A]+\frac{\partial A}{\partial t},
\end{equation}
and $\mathcal{L}^{*}[A]$ is the full Lindblad super operator as a
function of $A$. However, since $\rho^{S}$ still appears in eq.
\ref{KosGev} their definitions for heat flux and power \cite{Geva00}
are formally equivalent with Alicki's definitions in the
Schr\"{o}dinger picture:
\begin{eqnarray}
\dot{Q}&=&\textrm{Tr}\{\rho^{S}\mathcal{L}_{d}^{*}[H]\}=\textrm{Tr}\{\mathcal{L}_{d}[\rho^{S}]H^{S}\}\nonumber\\
P&=&\textrm{Tr}\left\{\rho^{S}\left(\frac{\partial H}{\partial
t}\right)^{S}\right\},\label{QPHeis}
\end{eqnarray}
where $\mathcal{L}_{d}^{*}[H]$ is the dissipative Lindblad super
operator as a function of $H$.

\subsubsection{The interaction picture\label{sec:Interaction}}

Consider the following Hamiltonian:
\begin{equation}
H(t)=H_{0}+V(t),\label{intH}
\end{equation}
where $H_{0}$ is usually a simple Hamiltonian (for example the
eigenenergy Hamiltonian of a system), and $V(t)$ is the coupling to
some external degree of freedom or field. The interaction
representation is defined by:
\begin{equation}
A^{I}=U_{0}^{\dag}(t)A^{S}U_{0}(t),\label{Interaction}
\end{equation}
where $U_{0}$ is the propagator associated with $H_{0}$. The average
value of an operator is independent of any such transformation:
\begin{equation}
\langle
A^{S}\rangle=\textrm{Tr}\{\rho^{S}A^{S}\}=\textrm{Tr}\{U_{0}^{\dag}(t)\rho^{S}U_{0}(t)U_{0}^{\dag}(t)A^{S}U_{0}(t)\}=\textrm{Tr}\{\rho^{I}A^{I}\}=\langle
A^{I}\rangle.
\end{equation}
Differentiation of $\langle A^{I}\rangle$ yields:
\begin{equation}
\frac{d\langle
A^{I}\rangle}{dt}=\frac{d(\textrm{Tr}\{\rho^{I}A^{I}\})}{dt}=\textrm{Tr}\left\{\frac{d\rho^{I}}{dt}A^{I}\right\}+\textrm{Tr}\left\{\rho^{I}\frac{dA^{I}}{dt}\right\}.\label{Int}
\end{equation}
Substituting eq. \ref{Interaction} into the two terms in eq.
\ref{Int} and using the relation
$\dot{U}_{0}=-\frac{i}{\hbar}H_{0}U_{0}$ yields:
%\begin{equation}\hspace{73mm}a = b \label{nnnn}\hspace{73mm}\mbox{(\ref{nnnn}a)}\nonumber\end{equation}
\begin{eqnarray}
\textrm{Tr}\left\{\frac{d\rho^{I}}{dt}A^{I}\right\}&=&-\frac{i}{\hbar}\textrm{Tr}\{\rho^{I}[A^{I},V^{I}]\}+\textrm{Tr}\{\mathcal{L}_{d}^{I}[\rho^{I}]A^{I}\}\label{Intsplita}\\
\textrm{Tr}\left\{\rho^{I}\frac{dA^{I}}{dt}\right\}&=&\frac{i}{\hbar}\textrm{Tr}\{\rho^{I}[H_{0}^{I},A^{I}]\}+\textrm{Tr}\left\{\rho^{I}\left(\frac{\partial
A}{\partial t}\right)^{I}\right\},\label{Intsplitb}
\end{eqnarray}
where $\mathcal{L}_{d}^{I}[\rho^{I}]$ is the dissipative Lindblad
super operator in the interaction picture. Substituting $H^{I}$ into
eq. \ref{Intsplita} and eq. \ref{Intsplitb} yields:
\begin{eqnarray}
\textrm{Tr}\left\{\frac{d\rho^{I}}{dt}H^{I}\right\}\!\!\!&=&\!\!\!-\frac{i}{\hbar}\textrm{Tr}\{\rho^{I}\![H_{0}^{I},V^{I}]\}\!\!+\!\!\textrm{Tr}\{\mathcal{L}_{d}^{I}[\rho^{I}]H^{I}\}
\!\!=\!\!-\frac{i}{\hbar}\textrm{Tr}\{\rho^{S}\![H_{0}^{S},V^{S}]\}\!\!+\!\!\textrm{Tr}\{\mathcal{L}_{d}[\rho^{S}]H^{S}\}\label{IntsplitHa}\\
\textrm{Tr}\left\{\rho^{I}\frac{dH^{I}}{dt}\right\}\!\!&=&\!\!\frac{i}{\hbar}\textrm{Tr}\{\rho^{I}[H_{0}^{I},\!\!V^{I}]\}\!\!+\!\!\textrm{Tr}\left\{\!\!\rho^{I}\!\!\left(\frac{\partial
H}{\partial
t}\right)^{I}\right\}\!\!=\!\frac{i}{\hbar}\textrm{Tr}\{\rho^{S}[H_{0}^{S},\!\!V^{S}]\}\!\!+\!\!\textrm{Tr}\left\{\!\rho^{S}\!\!\left(\!\!\frac{\partial
H}{\partial t}\!\right)^{S}\right\}\!\!.\label{IntsplitHb}
\end{eqnarray}
Adding eq. \ref{IntsplitHa} and eq. \ref{IntsplitHb} yields $d
E^{S}/dt$. However, each term differs from the corresponding terms
in the Schr\"{o}dinger picture by $\mp i/\hbar
\textrm{Tr}\{\rho^{S}[H_{0}^{S},V^{S}]\}$ (see eq. \ref{QSch} and
eq. \ref{PSch}). Thus heat and work should be redefined to avoid a
consistency problem when one moves from the Schr\"{o}dinger picture
to the interaction picture. We redefine heat and work as follows:
\begin{eqnarray}
\dot{Q}=
\textrm{Tr}\left\{\frac{\partial\rho(t')}{\partial t'}H(t')\right\}\nonumber\\
P=\textrm{Tr}\left\{\rho(t')\frac{\partial H(t')}{\partial
t'}\right\}.\label{QWB}
\end{eqnarray}
The definitions for heat and work in eq. \ref{QWB} differ from
Alicki's definitions for heat and work (eq. \ref{heatAL} and eq.
\ref{workAL} respectively) only by replacing the full derivatives
with respect to time with the partial derivatives with respect to
time. This replacement is necessary as it avoids the extra
commutator term $\mp i/\hbar
\textrm{Tr}\{\rho^{S}[H_{0}^{S},V^{S}]\}$ which arises from the full
time derivative in eq. \ref{Int}.

\subsection{Alternative approach to heat and work for unipartite
systems coupled to a time dependent external field\label{sec:Alt}}

Consider a unipartite system that is governed by eq. \ref{Master},
and is driven by a time dependent Hamiltonian of the form of eq.
\ref{intH}. The energy flux of such a system is given by
substitution of $H(t)$ into eq. \ref{Sch} in the Schr\"{o}dinger
picture:
\begin{equation}
\dot{E}=\frac{d\langle
H^{S}\rangle}{dt}=\textrm{Tr}\left\{\frac{\partial\rho^{S}}{\partial
t}H^{S}\right\}+\textrm{Tr}\left\{\rho^{S}\frac{\partial
H^{S}}{\partial t}\right\}.\label{SchS}
\end{equation}
The average value of the Hamiltonian ($\langle
E\rangle=\textrm{Tr}\{\rho^{S}H^{S}(t)\}$) contains contributions
from both the system energy and the interaction energy. Even if the
system reaches a steady state ($\dot{\rho}=0$), the quantity
$\dot{E}$ is not necessarily zero at steady state due to the term
$\textrm{Tr}\left\{\rho^{S}\frac{\partial H^{S}}{\partial
t}\right\}$. Let us consider the energy flux of the system alone
($\langle E_{0}\rangle=\textrm{Tr}\{\rho^{S}H_{0}^{S}(t)\}$):
\begin{equation}
\dot{E_{0}}=\frac{d\langle
H_{0}^{S}\rangle}{dt}=\textrm{Tr}\left\{\frac{\partial\rho^{S}}{\partial
t}H^{S}_{0}\right\}=\textrm{Tr}\{\mathcal{L}_{d}[\rho^{S}]H_{0}^{S}\}+\textrm{Tr}\{\mathcal{L}_{h}[\rho^{S}]H_{0}^{S}\}.\label{SchAl}
\end{equation}
We now define heat flux by:
\begin{equation}
\dot{Q}_{0}\equiv\textrm{Tr}\{\mathcal{L}_{d}[\rho]H_{0}\},\label{dotQ0}
\end{equation}
and power by:
\begin{equation}
P_{0}\equiv\textrm{Tr}\{\mathcal{L}_{h}[\rho]H_{0}\}=-\frac{i}{\hbar}\textrm{Tr}\{\rho[H_{0},V(t)]\}.\label{P0}
\end{equation}
These definitions are novel to the best of our knowledge, and have
the property of being identical in all physical pictures (the
expression in the Heisenberg and interaction pictures can be
obtained easily through the cyclic invariance of the trace). This
partitioning of the energy flux captures the spirit of Alicki's
partitioning in the sense that power is defined via the Hamiltonian
superoperator and heat flux is defined via the dissipative Lindblad
superoperator (although the present definition for $\dot{Q}_{0}$
involves only the bare Hamiltonian $H_{0}$ whereas Alicki's
definition for $\dot{Q}$ involves the full Hamiltonian
$H=H_{0}+V(t)$). Moreover, this partitioning of the energy flux has
the same structure as the partitioning we will use in section IIIA
(eq. \ref{WAA} and eq. \ref{QAA}) where the energy of a partially
traced system within a bipartite system is considered. Finally, note
that the quantity $\langle E_{0}\rangle$ does reach a steady state
since $H_{0}$ is time independent.

\subsection{The second law for unipartite systems\label{sec:SecondUni}}

The ultimate test for partitioning energy should be the fulfillment
of the second law. The most general formulation of the second law is
that of Clausius, which states that the total entropy change of a
closed system must be greater than or equal to zero. In that
respect, an open system together with the reservoirs that are
coupled to it is in fact closed. Thus the entropy changes of the
system and of the reservoirs must fulfill the relation: $\Delta
S_{total}=\Delta S_{system}+\Delta S_{reservoirs}\geq0$. For this
relation to hold there is no need to define work. When the system
operates in a cycle, the only contribution to the total entropy
change is the entropy change of the reservoirs ($\Delta
S_{system}=0$), and one can derive a Carnot formulation of the
second law, which sets a limit on the engine efficiency. However, in
order to have a Carnot formulation of the second law, work must be
defined.

To derive the Clausius and Carnot formulations of the
non-equilibrium second law of thermodynamics in differential form,
one replaces the equilibrium thermodynamic quantities of heat and
work by thermodynamic currents (or fluxes) of heat flux and power.
In order to satisfy Clausius's formulation of the second law for
driven unipartite systems \textit{at all times} we believe that one
has to use Alicki's definition for heat flux. However, at steady
state, in order to obtain Carnot's formula one must use our
definitions (especially for power), as we now explain.

We begin with Spohn's entropy production function \cite{Spohn} for a
system coupled to two reservoirs:
\begin{equation}
\sigma=\frac{\partial S}{\partial
t}-\frac{\dot{Q}_{H}}{T_{H}}-\frac{\dot{Q}_{C}}{T_{C}},\label{SpohnCarnot}
\end{equation}
where $\frac{\partial S}{\partial t}$ is the system's entropy change
defined via the von Neumann entropy, and
$-\frac{\dot{Q}_{H(C)}}{T_{H(C)}}$ is the reversible entropy current
from/to the hot (cold) thermal reservoir. Spohn showed that for a
completely positive map (such as the Lindblad super operator)
\cite{Spohn}:
\begin{equation}
\sigma\geq0.\label{entprod1}
\end{equation}
Eq. \ref{entprod1} is Clausius's general formulation of the second
law in differential form, which is valid at all times.

At steady state $\frac{\partial S}{\partial t}=0$, and hence we need
to analyze only the thermodynamic currents from/to the reservoirs.
Note that $\dot{Q}_{H(C)}=\textrm{Tr}\{\mathcal{L}_{dH(C)}[\rho]H\}$
has two contributions, namely
$\dot{Q}_{H(C)}=\dot{Q}_{H(C)0}+\dot{Q}_{H(C)V}$, where
$\dot{Q}_{H(C)V}=\textrm{Tr}\{\mathcal{L}_{dH(C)}[\rho]V\}$. Writing
eq. \ref{SchAl} for the case of two reservoirs yields:
\begin{equation}
\dot{E}_{0}=\dot{Q}_{H0}+\dot{Q}_{C0}+P_{0}.\label{SchHdev}
\end{equation}
At steady state ($\dot{E}_{0}=0$),
$\dot{Q}_{C0}^{ss}=-(\dot{Q}_{H0}^{ss}+P_{0}^{ss})$. If we assume
that $\dot{Q}_{C(H)V}^{ss}=0$, substitution of eq. \ref{SchHdev}
into eq. \ref{SpohnCarnot} yields Carnot's famous formula in
differential form:
\begin{equation}
\eta\equiv-\frac{P_{0}}{\dot{Q}_{H0}}\leq\frac{T_{H}-T_{C}}{T_{H}}.
\end{equation}
Note that a similar Carnot formulation of the second law does not
generally exist using Alicki's formalism, due to the fact that
generally $\dot{E}\neq0$, for example in the non-resonant
semiclassical ED-JCM discussed in \cite{Erez03}.

For completeness, we note that if $\dot{Q}_{C(H)V}=0$, eq.
\ref{SpohnCarnot} is trivially satisfied.

\section{Thermodynamics of bipartite systems with time independent
Hamiltonians\label{sec:Bipartite}}

\subsection{Energy fluxes for bipartite systems\label{sec:Bipartitea}}

Consider a bipartite system whose evolution is governed by eq.
\ref{Master}. The time independent bipartite Hamiltonian operates in
a $C^{m}\otimes C^{n}$ Hilbert space, and it is given by:
\begin{equation}
\mbox{\boldmath$H$}=\mbox{\boldmath$H_{A}$}+\mbox{\boldmath$H_{B}$}+\mbox{\boldmath$V_{AB}$},
\end{equation}
where $\mbox{\boldmath$H_{A}$}=H_{A}\otimes\openone_{B}$ and
$\mbox{\boldmath$H_{B}$}=\openone_{A}\otimes H_{B}$. For simplicity
we consider direct dissipation through subsystem $A$ only. However,
the analysis that follows can be generalized easily to the case
where both subsystems dissipate energy directly to the environment.
The generic form of the dissipative Lindblad super operator is given
by:
\begin{equation}
\mathcal{L}_{d}[\mbox{\boldmath$\rho_{AB}$}]=\sum_{k}\Gamma_{k}\{a_{k}([\mbox{\boldmath$\sigma_{k}^{\dag}$},\mbox{\boldmath$\sigma_{k}\rho_{AB}$}]+[\mbox{\boldmath$\rho_{AB}\sigma_{k}^{\dag}$},\mbox{\boldmath$\sigma_{k}$}])+b_{k}([\mbox{\boldmath$\sigma_{k}$},\mbox{\boldmath$\sigma_{k}^{\dag}\rho_{AB}$}]+[\mbox{\boldmath$\rho_{AB}\sigma_{k}$},\mbox{\boldmath$\sigma_{k}^{\dag}$}])\}
\equiv\sum_{k}\mathcal{L}_{d}^{(k)}[\mbox{\boldmath$\rho_{AB}$}],
\end{equation}
where $\mbox{\boldmath$\sigma_{k}$}
(\mbox{\boldmath$\sigma_{k}^{\dag}$})$ is the $k$th lowering
(raising) operator of subsystem $A$ in the tensor product space
($\mbox{\boldmath$\sigma_{k}$}
(\mbox{\boldmath$\sigma_{k}^{\dag}$})=\sigma_{kA}(\sigma_{kA}^{\dag})\otimes\openone_{B}$),
$\Gamma_{k}$ is the $k$th decay rate, $a_{k}$, and $b_{k}$ are
prefactors whose ratio is a function of the temperature of the $k$th
reservoir.

The derivation that follows is general for either the
Schr\"{o}dinger, Heisenberg, or interaction representation. For
convenience we will work in the Schr\"{o}dinger representation.
Using the definition for heat flux (eq. \ref{QWB}), the energy flux
of the full bipartite system is given by:
\begin{equation}
\dot{E}_{AB}=\textrm{Tr}\{\mathcal{L}_{d}[\mbox{\boldmath$\rho_{AB}$}]\mbox{\boldmath$H$}\}\equiv\dot{Q}=\sum_{k}\dot{Q}_{k},\label{ABflux}
\end{equation}
where
$\dot{Q}_{k}=\textrm{Tr}\{\mathcal{L}_{d}^{(k)}\mbox{\boldmath$H$}\}$
is the heat flux associated with coupling to the $k$th reservoir.
Note that since the Hamiltonian is time independent there is no work
involved. Thus, to an outside observer looking on the bipartite
system as a whole the full system is only dissipating. Is this the
whole story? No.

Let us examine the energy flux of subsystem $A$ (the subsystem for
which direct dissipation occurs):
\begin{equation}
\dot{E}_{A}\equiv\textrm{Tr}_{A}\{\dot{\rho}_{A}H_{A}\}=
\textrm{Tr}\{\mathcal{L}[\mbox{\boldmath$\rho_{AB}$}]\mbox{\boldmath$H_{A}$}\}=-\frac{i}{\hbar}\textrm{Tr}\{\mbox{\boldmath$\rho_{AB}$}[\mbox{\boldmath$H_{A}$},\mbox{\boldmath$V_{AB}$}]\}+\textrm{Tr}\{\mathcal{L}_{d}[\mbox{\boldmath$\rho_{AB}$}]\mbox{\boldmath$H_{A}$}\},\label{EdotA}
\end{equation}
where $H_{A}$ is the Hamiltonian of subsystem $A$ without the tensor
product with $\openone_{B}$, and
$\dot{\rho}_{A}=\textrm{Tr}_{B}\{\mbox{\boldmath$\rho_{AB}$}\}$ is
the partial density matrix of subsystem $A$. The first equality in
equation \ref{EdotA} stems from the following algebra:
\begin{eqnarray}
\dot{E}_{A}&=&\textrm{Tr}_{A}\{\dot{\rho}_{A}H_{A}\}=\sum_{i,j}{\dot{\rho}_{A}^{ij}H_{A}^{ji}}=\sum_{i,j}{(\sum_{\alpha}{\mbox{\boldmath$\dot{\rho}_{AB}^{i\alpha,j\alpha}$}})H_{A}^{ji}}\nonumber\\
&=&
\!\!\!\!\!\sum_{i,j,\alpha,\beta}{\mbox{\boldmath$\dot{\rho}_{AB}^{i\alpha,j\beta}$}H_{A}^{ji}\delta_{B}^{\beta\alpha}}=
\sum_{i,j,\alpha,\beta}{\mbox{\boldmath$\dot{\rho}_{AB}^{i\alpha,j\beta}H_{A}^{j\beta,i\alpha}$}}=\textrm{Tr}_{A,B}\{\mbox{\boldmath$\dot{\rho}_{AB}H_{A}$}\}=\textrm{Tr}\{\mathcal{L}[\mbox{\boldmath$\rho_{AB}$}]\mbox{\boldmath$H_{A}$}\}\label{algebricex}.
\end{eqnarray}
Comparing eq. \ref{EdotA} and eq. \ref{algebricex} we associate the
first term on the RHS of eq. \ref{EdotA} with power:
\begin{equation}
P_{A}\equiv-\frac{i}{\hbar}\textrm{Tr}\{\mbox{\boldmath$\rho_{AB}$}[\mbox{\boldmath$H_{A}$},\mbox{\boldmath$V_{AB}$}]\},\label{WAA}
\end{equation}
while we associate the second term with heat flux:
\begin{equation}
\dot{Q}_{A}\equiv\textrm{Tr}\{\mathcal{L}_{d}[\mbox{\boldmath$\rho_{AB}$}]\mbox{\boldmath$H_{A}$}\}.\label{QAA}
\end{equation}
Before we continue we wish to emphasize the differences between our
novel definitions for heat flux and power (eq. \ref{QAA} and eq.
\ref{WAA}, respectively) and these of Alicki (eq. \ref{heatAL} and
eq. \ref{workAL}, respectively). Firstly, our definitions refer to a
single degree of freedom within a bipartite system. Secondly, the
dimensionality of the operators in our definitions is $m\otimes n$
(due to the expansion in eq. \ref{algebricex}) as opposed to $m$
when a single degree of freedom is coupled to an external field
(Alicki's work). Thirdly, power emerges simply from the commutator
between $H_{A}$ and $V_{AB}$, and not from a time dependent field.
The last difference is striking since it indicates that the
Hamiltonian need not depend on time to get work. In fact, whenever
an external field is replaced by a quantized degree of freedom, the
time dependence of the Hamiltonian is removed at the expense of
increasing the dimensionality of the Hamiltonian.

Similarly, we can analyze the energy flux of subsystem $B$:
\begin{equation}
\dot{E}_{B}\equiv \textrm{Tr}_{B}\{\dot{\rho}_{B}H_{B}\}=
-\frac{i}{\hbar}\textrm{Tr}\{\mbox{\boldmath$\rho_{AB}$}[\mbox{\boldmath$H_{B}$},\mbox{\boldmath$V_{AB}$}]\}=P_{B}.\label{EdotB}
\end{equation}
There is no contribution to the energy flux of subsystem $B$ from
the dissipative part of the Lindblad super operator (this is
physically expected since the dissipation is through subsystem $A$
only):
\begin{eqnarray}
\textrm{Tr}\{\mathcal{L}_{d}[\mbox{\boldmath$\rho_{AB}$}]\mbox{\boldmath$H_{B}$}\}&=&-\sum_{k}\Gamma_{k}\textrm{Tr}\left\{(a_{k}([\mbox{\boldmath$\sigma_{k}^{\dag}$},\mbox{\boldmath$\sigma_{k}\rho_{AB}$}]+[\mbox{\boldmath$\rho_{AB}\sigma_{k}^{\dag}$},\mbox{\boldmath$\sigma_{k}$}])\right.\nonumber\\
&+&\left.b_{k}([\mbox{\boldmath$\sigma_{k}$},\mbox{\boldmath$\sigma_{k}^{\dag}\rho_{AB}$}]+[\mbox{\boldmath$\rho_{AB}\sigma_{k}$},\mbox{\boldmath$\sigma_{k}^{\dag}$}]))\mbox{\boldmath$H_{B}$}\right\}=0.\label{zeroheatB}
\end{eqnarray}
To see this consider just the terms
$\textrm{Tr}\{([\mbox{\boldmath$\sigma_{k}^{\dag}$},
\mbox{\boldmath$\sigma_{k}\rho_{AB}$}]+[\mbox{\boldmath$\rho_{AB}\sigma_{k}^{\dag}$},
\mbox{\boldmath$\sigma_{k}$}])\mbox{\boldmath$H_{B}$}\}$:
\begin{eqnarray*}
\textrm{Tr}\{([\mbox{\boldmath$\sigma_{k}^{\dag}$}, \mbox{\boldmath$\sigma_{k}\rho_{AB}$}]\!\!&+&\!\![\mbox{\boldmath$\rho_{AB}\sigma_{k}^{\dag}$}, \mbox{\boldmath$\sigma_{k}$}])\mbox{\boldmath$H_{B}$}\}=\textrm{Tr}\{\mbox{\boldmath$\rho_{AB}$}(\mbox{\boldmath$H_{B}\sigma_{k}^{\dag}\sigma_{k}$}-2\mbox{\boldmath$\sigma_{k}^{\dag}H_{B}\sigma_{k}$}+\mbox{\boldmath$\sigma_{k}^{\dag}\sigma_{k}H_{B}$})\}\\
&=&\textrm{Tr}\{\mbox{\boldmath$\rho_{AB}$}[(\openone_{A}\otimes
H_{B})(\sigma_{kA}^{\dag}\otimes\openone_{B})(\sigma_{kA}\otimes\openone_{B})-2(\sigma_{kA}^{\dag}\otimes\openone_{B})(\openone_{A}\otimes
H_{B})(\sigma_{kA}\otimes\openone_{B})\\
&+&(\sigma_{kA}^{\dag}\otimes\openone_{B})(\sigma_{kA}\otimes\openone_{B})(\openone_{A}\otimes
H_{B})\}=(2-2)\textrm{Tr}\{\mbox{\boldmath$\rho_{AB}$}\sigma_{kA}^{\dag}\sigma_{A}\otimes
H_{B}\}=0.
\end{eqnarray*}
Similarly, $\textrm{Tr}\{([\mbox{\boldmath$\sigma_{k}$},
\mbox{\boldmath$\sigma_{k}^{\dag}\rho_{AB}$}]+[\mbox{\boldmath$\rho_{AB}\sigma_{k}$},
\mbox{\boldmath$\sigma_{k}^{\dag}$}])\mbox{\boldmath$H_{B}$}\}=0$.

The first law of thermodynamics in differential form can be
formulated in two different ways. The first way relies on looking at
each subsystem independently; for subsystem $A$ it is obtained by
combining eq. \ref{EdotA} with the definitions in eq. \ref{QAA} and
eq. \ref{WAA}:
\begin{equation}
\dot{E}_{A}=\dot{Q}_{A}+P_{A}.\label{EdotAfinal}
\end{equation}
The first law for subsystem $B$ is given by eq. \ref{EdotB}.

The second way relies on looking at the full bipartite system; it is
obtained by substituting eq. \ref{QAA} for the heat flux into eq.
\ref{ABflux}:
\begin{equation}
\dot{E}_{AB}=\dot{Q}_{AB}=\dot{Q}_{A}+\dot{Q}_{V},
\end{equation}
where $\dot{Q}_{V}\equiv
\textrm{Tr}\{\mathcal{L}_{d}[\mbox{\boldmath$\rho_{AB}$}]\mbox{\boldmath$V_{AB}$}\}$
and $\dot{Q}_{A}$ was defined in eq. \ref{QAA}. By adding together
eq. \ref{EdotAfinal} and eq. \ref{EdotB} we see that in general:
\begin{equation}
\dot{E}_{A}+\dot{E}_{B}=P_{A}+\dot{Q}_{A}+P_{B}\neq\dot{Q}_{A}+\dot{Q}_{V}=
\dot{E}_{AB}.
\end{equation}

\subsection{The second law for bipartite systems\label{sec:Bipartiteb}}

The second law of thermodynamics in differential form for the full
bipartite system is once more obtained via Spohn's entropy
production function defined in \cite{Spohn}, \cite{Alicki}:
\begin{equation}
\sigma\equiv\frac{\partial S_{AB}}{\partial
t}+J\geq0,\label{entprod0}
\end{equation}
where $\frac{\partial S_{AB}}{\partial t}$ is the entropy production
associated with the full bipartite density matrix via
differentiation of the von Neumann entropy \cite{von Neumann}
($S_{sys}=-k_{B}\textrm{Tr}\{\mbox{\boldmath$\rho_{AB}$}\ln{\mbox{\boldmath$\rho_{AB}$}}\}$),
and $J$ is the entropy production associated with the reservoirs
(via the heat flux from/to the reservoirs) and it is given by:
\begin{equation}
J=-\sum_{k}\beta_{k}\dot{Q}_{k}.
\end{equation}
Here $-\beta_{k}\dot{Q}_{k}$ is the differential change in the
entropy of the $k$th reservoir, and $\beta_{k}=1/k_{B}T_{k}$ where
$T_{k}$ is the temperature of the $k$th reservoir. Eq.
\ref{entprod0} represents the differential form of the second law of
thermodynamics in Clausius's formulation since the sum of the
entropy changes of the system and reservoirs is guaranteed to be
positive.

Is it possible to formulate a second law of thermodynamics for a
subsystem within a bipartite system? Consider an entropy production
function that is based on thermodynamic currents of subsystem $A$
only, which is coupled to two thermal reservoirs:
\begin{equation}
\sigma_{A}\equiv\frac{\partial S_{A}}{\partial
t}+J_{A},\label{entprodA}
\end{equation}
where $\frac{\partial S_{A}}{\partial t}$ is the entropy production
associated with the partial density matrix via differentiation of
the partial von Neumann entropy \cite{von Neumann}, and $J_{A}$ is
the entropy production associated with the \textit{direct} coupling
between subsystem $A$ and the reservoirs:
\begin{equation}
J_{A}\equiv-\beta_{H}\dot{Q}_{HA}-\beta_{C}\dot{Q}_{CA},
\end{equation}
where $\dot{Q}_{HA}\equiv\textrm{Tr}\{\mathcal{L}_{dH}[\rho]
\mbox{\boldmath$H_{A}$}\}$, and similarly for $\dot{Q}_{CA}$. Note
that each heat flux component in $J$ has two contributions, e.g.
$\dot{Q}_{H}=\dot{Q}_{HA}+\dot{Q}_{HV}$, where
$\dot{Q}_{HV}\equiv\textrm{Tr}\{\mathcal{L}_{dH}[\rho]
\mbox{\boldmath$V_{AB}$}\}$. Unlike $\sigma$, $\sigma_{A}$ is
\textit{not} necessarily positive due to the oscillatory nature of
the partial entropies under Hamiltonian evolution, for example, the
partial entropies in the Jaynes-Cummings model (JCM) \cite{Knight}
\cite{Erez01}. However, due to the dissipative nature of the
Lindblad superoperator subsystem $A$ is expected to reach a steady
state, at which point $\frac{\partial S_{A}}{\partial t}=0$ and
$\dot{E}_{A}=0$. Moreover, at steady state, (1) if
$\dot{Q}_{HV}=\dot{Q}_{CV}=0$ (which implies that $J=J_{A}$), and
(2) if the main source for entropy production is the dissipation of
subsystem $A$ ($J>\frac{\partial S_{AB}}{\partial t}$), then
\begin{equation}
\sigma_{A}^{ss}=J_{A}=J > 0.\label{SpohnA}
\end{equation}
If the above two conditions are fulfilled, eq. \ref{SpohnA}
corresponds to Clausius's formulation of the second law for the
subsystem $A$, and Carnot's formula follows from it trivially. Note
that condition (1) appeared in our earlier discussion of Spohn's
entropy production formula for unipartite systems (Section IV). Eq.
\ref{SpohnA} is verified numerically for the quantum ED-JCM in a
forthcoming publication \cite{Erez03}.

It is interesting to consider the differences between $\sigma_A$ for
the partially traced system and $\sigma$ for the forced unipartite
system described in Section \ref{sec:SecondUni}. We expect
deviations between these two to be significant when the entropy of
the subsystems oscillate, since the unipartite system entropy can
only increase. Thus, at short times we expect that there may be
significant differences, while at steady state, when the entropy of
the partially traced bipartite system becomes constant, we expect
the differences to be small. This is borne out by numerical studies,
where transient collapses and revivals are observed in the
partially-traced bipartite treatment at short times, while only
minor deviations from the unipartite values are observed at steady
state \cite{Erez03}. Note that these are the same oscillations that
are responsible for the breakdown of the Spohn entropy production
formula for the partially-traced bipartite system, as discussed in
Eq. \ref{SpohnA}.

\subsection{Applications\label{sec:Apps}}

\subsubsection{The Jaynes-Cummings model (JCM)\label{sec:JCM}}

The Jaynes-Cummings model is the simplest fully quantum model that
describes the interaction between light and matter \cite{JCM}. In
this model a two-level material system interacts resonantly with a
single quantized cavity mode according to the following Hamiltonian
(here and throughout the whole section we omit the superscript $S$):
\begin{equation}
\mbox{\boldmath$H$}=\mbox{\boldmath$H_{m}$}+\mbox{\boldmath$H_{f}$}+\mbox{\boldmath$V_{mf}$},\label{JCMH}
\end{equation}
where
$\mbox{\boldmath$H_{m}$}=\hbar\omega_{a}\sigma_{e}\otimes\openone_{f}$
is the matter Hamiltonian,
$\mbox{\boldmath$H_{f}$}=\hbar\omega_{f}\openone_{a}\otimes
a^{\dag}a$ is the field Hamiltonian, and
$\mbox{\boldmath$V_{mf}$}=\hbar\lambda[\sigma^{+}\otimes
a+\sigma^{-}\otimes a^{\dag}]$ is the coupling Hamiltonian
($\sigma^{+}$ and $\sigma^{-}$ are the raising and lowering matter
operators, respectively). The matrix form of the matter operators is
given by:
\begin{equation*}
\sigma^{+}=\left(
\begin{array}{cc}
0 & 1\\
0 & 0\\
\end{array}\right)\ \ \ \ \
\sigma^{-}=\left(
\begin{array}{cc}
0 & 0\\
1 & 0\\
\end{array}\right)\ \ \ \ \
\sigma_{e}=\left(
\begin{array}{cc}
1 & 0\\
0 & 0\\
\end{array}\right).
\end{equation*}
The system is governed by the following master equation in the
Schr\"{o}dinger picture:
\begin{equation}
\mbox{\boldmath$\dot{\rho}_{mf}$}=\mathcal{L}_{h}[\mbox{\boldmath$\rho_{mf}$}],
\end{equation}
where
$\mathcal{L}_{h}[\mbox{\boldmath$\rho_{mf}$}]=-\frac{i}{\hbar}[\mbox{\boldmath$H$},\mbox{\boldmath$\rho_{mf}$}]$
is the Hamiltonian superoperator. The energy flux of the full
matter-field system is given by:
\begin{equation}
\dot{E}_{mf}=\textrm{Tr}\{\mathcal{L}_{h}[\mbox{\boldmath$\rho_{mf}]H$}\}=0.\label{dotaf}
\end{equation}
Note that since the evolution is purely unitary the total
matter-field energy is constant. Let us examine the energy flux of
the matter:
\begin{equation}
\dot{E}_{m}=
\textrm{Tr}_{m}\{\dot{\rho}_{m}H_{m}\}=-\frac{i}{\hbar}\textrm{Tr}\{\mbox{\boldmath$\rho_{mf}$}[\mbox{\boldmath$H_{m}$},\mbox{\boldmath$V_{mf}$}]\}=P_{m},\label{dota}
\end{equation}
where $H_{m}$ is the Hamiltonian of the matter without the tensor
product with $\openone_{f}$, and $P_{m}$ is the power associated
with the atom. The energy flux of the selected cavity mode (field)
is given by:
\begin{equation}
\dot{E}_{f}=
\textrm{Tr}_{f}\{\dot{\rho}_{f}H_{f}\}=-\frac{i}{\hbar}\textrm{Tr}\{\mbox{\boldmath$\rho_{mf}$}[\mbox{\boldmath$H_{f}$},\mbox{\boldmath$V_{mf}$}]\}=P_{f},\label{dotf}
\end{equation}
where $H_{f}$ is the Hamiltonian of the field without the tensor
product with $\openone_{m}$, and $P_{f}$ is the power associated
with the field. Generally,
$\dot{E}_{m}+\dot{E}_{f}=\textrm{Tr}\{\mbox{\boldmath$\rho_{mf}$}[\mbox{\boldmath$H_{m}$}+\mbox{\boldmath$H_{f}$},\mbox{\boldmath$V_{mf}$}]\}=\hbar^{2}\lambda\textrm{Tr}\{\mbox{\boldmath$\rho_{mf}$}(\omega_{m}-\omega_{f})(\sigma^{+}\otimes
a-\sigma^{-}\otimes a^{\dag})\}\neq\dot{E}_{mf}=0$. However, under
perfect atomic-field resonance
($[\mbox{\boldmath$H_{m}$}+\mbox{\boldmath$H_{f}$},\mbox{\boldmath$V_{mf}$}]=0$):
\begin{equation}
P_{m}=-P_{f},
\end{equation}
and hence $\dot{E}_{m}+\dot{E}_{f}=\dot{E}_{mf}=0$.

No entropy is produced during the interaction:
\begin{equation}
\sigma=0.
\end{equation}
This can be understood in two ways. Firstly, under pure Hamiltonian
evolution the eigenvalues of the bipartite system are a constant of
the motion \cite{Erez01} and hence $\frac{\partial S}{\partial
t}=0$. Secondly, since there is no coupling to an external heat
reservoir (closed system) no heat is produced and hence $J=0$.
However, partial entropies do change with time and the entropy
content of the partial systems gives information with respect to
entanglement and disorder \cite{Erez01}.

\subsubsection{The extended dissipative
Jaynes-Cummings model (ED-JCM)\label{sec:ED-JCM}}

\begin{figure}[htb]
\begin{center}
\includegraphics{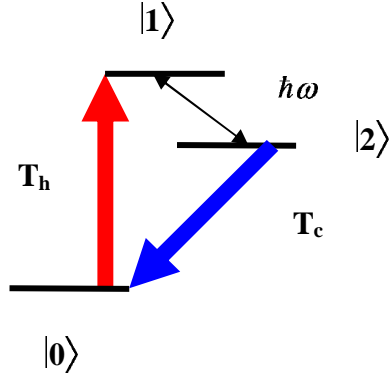}
\end{center}
\caption{\label{EDJCM}Three level system interacting with two heat
reservoirs (hot and cold) and a quantized cavity mode.}
\end{figure}

Consider a three-level system interacting resonantly with one
quantized cavity mode and two thermal photonic reservoirs as
depicted in Fig. \ref{EDJCM}.
The system is governed by the following master equation in the
Schr\"{o}dinger picture:
\begin{equation}
\mbox{\boldmath$\dot{\rho}_{mf}$}=\mathcal{L}_{h}[\mbox{\boldmath$\rho_{mf}$}]+\mathcal{L}_{dC}[\mbox{\boldmath$\rho_{mf}$}]+\mathcal{L}_{dH}[\mbox{\boldmath$\rho_{mf}$}].\label{MQE}
\end{equation}
The Hamiltonian superoperator is given by:
$\mathcal{L}_{h}[\mbox{\boldmath$\rho_{mf}$}]=-\frac{i}{\hbar}[\mbox{\boldmath$H_{s}$},\mbox{\boldmath$\rho_{mf}$}]$,
where $\mbox{\boldmath$H$}\equiv
\mbox{\boldmath$H_{s}$}=\mbox{\boldmath$H_{m}$}+\mbox{\boldmath$H_{f}$}+\mbox{\boldmath$V_{mf}$}$
is a JCM type Hamiltonian,
$\mbox{\boldmath$H_{m}$}=\hbar\bar{\sigma}\otimes\openone_{f};\
H_{f}=\hbar\omega\openone_{a}\otimes a^{\dag}a;\
V_{af}=\hbar\lambda[\sigma_{21}\otimes
a^{\dag}+\sigma_{21}^{\dag}\otimes a]$.
$\mathcal{L}_{dC}[\mbox{\boldmath$\rho_{mf}$}]$ and
$\mathcal{L}_{dH}[\mbox{\boldmath$\rho_{mf}$}]$ are the dissipative
cold and hot Lindblad super operators, respectively:
\begin{eqnarray}
\!\!\!\!\!\!\!\!\!\!\!\!\!\!\!\!\!\!\!\!\!\!\!\!\!\mathcal{L}_{dC}[\mbox{\boldmath$\rho_{mf}$}]\!\!&=&\!\!\Gamma_{02}\{(n_{02}\!+\!1)([\mbox{\boldmath$\sigma_{02}\rho_{mf}$},\!\mbox{\boldmath$\sigma_{02}^{\dag}$}]\!+\![\mbox{\boldmath$\sigma_{02}$},\!\mbox{\boldmath$\rho_{mf}\sigma_{02}^{\dag}$}])\!+\!
n_{02}([\mbox{\boldmath$\sigma_{02}^{\dag}\rho_{mf}$},\!\mbox{\boldmath$\sigma_{02}$}]\!+\![\mbox{\boldmath$\sigma_{02}^{\dag}$},\!\mbox{\boldmath$\rho_{mf}\sigma_{02}$}])\}\nonumber\\
\!\!\!\!\!\!\!\!\!\!\!\!\!\!\!\!\!\!\!\!\!\!\!\!\!\mathcal{L}_{dH}[\mbox{\boldmath$\rho_{mf}$}]\!\!&=&\!\!\Gamma_{01}\{(n_{01}\!+\!1)([\mbox{\boldmath$\sigma_{01}\rho_{mf}$},\!\mbox{\boldmath$\sigma_{01}^{\dag}$}]\!+\![\mbox{\boldmath$\sigma_{01}$},\!\mbox{\boldmath$\rho_{mf}\sigma_{01}^{\dag}$}])\!+\!
n_{01}([\mbox{\boldmath$\sigma_{01}^{\dag}\rho_{mf}$},\!\mbox{\boldmath$\sigma_{01}$}]\!+\![\mbox{\boldmath$\sigma_{01}^{\dag}$},\!\mbox{\boldmath$\rho_{mf}\sigma_{01}$}])\},\label{LdCH}
\end{eqnarray}
where $\Gamma_{02}$ and $\Gamma_{01}$ are the Weiskopf-Wigner decay
constant associated with the cold and hot reservoirs, respectively,
and $n_{02}$ and $n_{01}$ are the number of thermal photons in the
cold and hot reservoirs, respectively. The temperature of the
thermal photonic reservoirs is given by:
\begin{equation}
T_{01(02)}=T_{H(C)}=\frac{\hbar(\omega_{1(2)}-\omega_{0})}{k_{B}\ln(1/n_{02(01)}+1)}\label{temperature},
\end{equation}
The matrix form of the matter operators is given by:
\begin{equation*}
\sigma_{21}=\left(
\begin{array}{ccc}
0 & 0 & 0\\
0 & 0 & 0\\
0 & 1 & 0
\end{array}\right)\ \ \ \ \
\sigma_{01}=\left(
\begin{array}{ccc}
0 & 1 & 0\\
0 & 0 & 0\\
0 & 0 & 0
\end{array}\right)\ \ \ \ \
\sigma_{02}=\left(
\begin{array}{ccc}
0 & 0 & 1\\
0 & 0 & 0\\
0 & 0 & 0
\end{array}\right)\ \ \ \ \
\bar{\sigma}=\left(
\begin{array}{ccc}
\omega_{0} & 0 & 0\\
0 & \omega_{1} & 0\\
0 & 0 & \omega_{2}
\end{array}\right).
\end{equation*}
The extended dissipative JCM master equation (eq. \ref{MQE}) can be
obtained by summing the Hamiltonian contribution and the two
dissipative contributions. Alternatively, it can be derived in a
similar fashion to the dissipative JCM master equation
\cite{Erez03}. Assuming that the nature of the interaction between
each pair of matter levels is dipole coupling, due to parity
considerations not all three transitions would be allowed by dipole
coupling. However, this issue is avoided in systems with a break in
symmetry \cite{Shapiro}.

Let us examine the energy flux of the matter:
\begin{equation}
\dot{E}_{m}=
\textrm{Tr}_{m}\{\dot{\rho}_{m}H_{m}\}=-\frac{i}{\hbar}\textrm{Tr}\{\mbox{\boldmath$\rho_{mf}$}[\mbox{\boldmath$H_{m}$},\mbox{\boldmath$V_{mf}$}]\}+\textrm{Tr}\{\mathcal{L}_{d}[\mbox{\boldmath$\rho_{mf}$}]\mbox{\boldmath$H_{m}$}\},\label{Edotm}
\end{equation}
where $H_{m}$ is the matter Hamiltonian without the tensor product
with $\openone_{f}$. By substitution into eq. \ref{EdotA}, heat flux
and power for the matter are given by:
\begin{eqnarray}
\dot{Q}_{m}&=&\textrm{Tr}\{\mathcal{L}_{dC}[\mbox{\boldmath$\rho_{mf}$}]\mbox{\boldmath$H_{m}$}\}+\textrm{Tr}\{\mathcal{L}_{dH}[\mbox{\boldmath$\rho_{mf}$}]\mbox{\boldmath$H_{m}$}\}=\dot{Q}_{mC}+\dot{Q}_{mH}\nonumber\\
P_{m}&=&-\frac{i}{\hbar}\textrm{Tr}\{\mbox{\boldmath$\rho_{mf}$}[\mbox{\boldmath$H_{m}$},\mbox{\boldmath$V_{mf}$}]\},\label{QWBm}
\end{eqnarray}
where
$\dot{Q}_{mC(H)}\equiv\textrm{Tr}\{\mathcal{L}_{dC(H)}[\mbox{\boldmath$\rho_{mf}$}]\mbox{\boldmath$H_{m}$}\}$.
The energy flux of the selected cavity mode (field):
\begin{equation}
\dot{E}_{f}=
\textrm{Tr}_{f}\{\dot{\rho}_{f}H_{f}\}=-\frac{i}{\hbar}\textrm{Tr}\{\mbox{\boldmath$\rho_{mf}$}[\mbox{\boldmath$H_{f}$},\mbox{\boldmath$V_{mf}$}]\}=P_{f},\label{Edotf}
\end{equation}
where $H_{f}$ is the field Hamiltonian without the tensor product
with $\openone_{m}$, and$P_{f}$ is the power associated with the
field.

The energy flux of the full matter-field system is given by:
\begin{equation}
\dot{E}_{mf}=\textrm{Tr}\{\mathcal{L}_{dC}[\mbox{\boldmath$\rho_{mf}$}]\mbox{\boldmath$H$}\}+\textrm{Tr}\{\mathcal{L}_{dH}[\mbox{\boldmath$\rho_{mf}$}]\mbox{\boldmath$H$}\},\label{JCMfluxaf}
\end{equation}
where each heat flux component is a sum of two contributions:
\begin{equation}
\textrm{Tr}\{\mathcal{L}_{dC(H)}[\mbox{\boldmath$\rho_{mf}$}]H\}=\textrm{Tr}\{\mathcal{L}_{dC(H)}[\mbox{\boldmath$\rho_{mf}$}](\mbox{\boldmath$H_{m}$}+\mbox{\boldmath$V_{mf}$})\}=\dot{Q}_{mC(H)}+\dot{Q}_{VC(H)},
\end{equation}
where
$\dot{Q}_{VC(H)}\equiv\textrm{Tr}\{\mathcal{L}_{dC(H)}[\mbox{\boldmath$\rho_{mf}$}]\mbox{\boldmath$V$}\}$.
Since we are in perfect matter-field resonance, $P_{m}=-P_{f}$
($[\mbox{\boldmath$H_{m}$},\mbox{\boldmath$V_{mf}$}]=-[\mbox{\boldmath$H_{f}$},\mbox{\boldmath$V_{mf}$}]$),
hence:
\begin{equation}
\dot{E}_{m}+\dot{E}_{f}=\dot{E}_{mf}-\dot{Q}_{V}.\label{fluxsum}
\end{equation}
$\dot{Q}_{V}$ vanishes if the off-diagonal matrix elements of
$\mbox{\boldmath$\rho_{mf}$}$ are purely imaginary. Note that to an
observer looking on the matter alone, work flux (power) and heat
fluxes correspond to the traditional view of the first law of
thermodynamics in which energy is divided into work and heat. The
field which is the work source either receives or emits energy to
the working medium (the matter) in the form of power.

In a forthcoming publication we give a full dynamical and
thermodynamical analysis of the ED-JCM, and show that it acts as a
quantum optical amplifier \cite{Erez03}.

\section{Conclusion\label{sec:Conc}}
\label{sec:conclusion}

 In this paper we have considered the thermodynamics of
unipartite systems (systems coupled to an external time dependent
field and heat reservoirs), as well as bipartite systems. In the
latter case, a supplementary part of the system replaces the time
dependent field.

For unpartite systems, we gave a minor generalization of Alicki's
definitions for heat flux and power, by extending these definitions
from the Schr\"{o}dinger to the Heisenberg and interaction pictures.
In this generalization, partial time derivatives replace full time
derivatives in the definitions for heat flux and power. We then
presented an alternative approach to the partitioning of the energy
flux into heat flux and power in which the interaction energy is not
included directly, that presages our bipartite treatment. We showed
that at steady state, if the interaction energy contribution to the
heat flux vanishes, this alternative partitioning leads to a Carnot
formulation of the second law.

We then turned to bipartite systems. We presented a novel definition
for power, based on the energy fluxes of the individual subsystems,
that is the natural generalization of our alternative partitioning
of energy flux in unipartite systems. The first law of
thermodynamics was derived in differential form in two different
ways. The first relies on looking at each subsystem independently,
while the second relies on looking at the full bipartite system.
Although any partitioning of the energy flux is consistent with the
first law, the ultimate test of a good partitioning is the
fulfillment of the second law. For the partitioning into (full
bipartite system) + (reservoirs) the second law in the Clausius form
follows almost trivially from Spohn's entropy production function.
However, for the partitioning into (subsystem A) + (subsystem B +
reservoirs) we found that there is generally no second law of the
Clausius type due to oscillations in the partial entropies of the
subsystems. Nevertheless, at steady state both the Clausius and
Carnot versions of the second law are satisfied under some
well-controlled conditions. In a forthcoming publication we show,
both analytically and numerically, that for the ED-JCM both the
Clausius and a Carnot versions of the second law are satisfied at
steady state \cite{Erez03}.

\section*{Acknowledgments}
We thank Prof. Eitan Geva for helpful comments regarding the
Heisenberg representation. This work was supported by the
German-Israeli Foundation for Scientific Research and Development.

\end{document}